\documentclass[aps,prb,twocolumn,showpacs,tightenlines,amsmath]{revtex4}
 \usepackage{graphics}
 \usepackage{epsfig}
   \begin{document}
   \title{\Large \bf{
Magnetic field symmetry of pump currents of adiabatically driven 
mesoscopic structures
         }}
\author{
M. Moskalets$^{1}$
and
M. B\"uttiker$^2$
}
\affiliation{
     $^1$Department of Metal and Semiconductor Physics,\\
     National Technical University "Kharkiv Polytechnic Institute",
     61002 Kharkiv, Ukraine\\
     $^2$D\'epartement de Physique Th\'eorique, Universit\'e de Gen\`eve,
     CH-1211 Gen\`eve 4, Switzerland\\}
\date\today
   \begin{abstract}
We examine the scattering properties of a slowly and periodically
driven mesoscopic sample using the Floquet function approach.
One might expect that at sufficiently low driving frequencies 
it is only the frozen scattering matrix which is important. 
The frozen scattering matrix reflects the properties
of the sample at a given instant of time.  Indeed many aspects of 
adiabatic scattering can be described in terms of the frozen scattering matrix. However, we demonstrate
that the Floquet scattering matrix, to first order in the driving frequency, 
is determined by an additional matrix which reflects the fact 
that the scatterer is time-dependent. This low frequency irreducible 
part of the Floquet matrix 
has symmetry properties with respect to
time and/or a magnetic field direction reversal opposite to
that of the frozen scattering matrix. 
We investigate the quantum rectification properties
of a pump
which additionally is subject to an external dc voltage.
We split the dc current flowing through the pump into several parts
with well defined properties with respect to a magnetic field and/or
an applied voltage inversion.
   \end{abstract}
\maketitle
\small

\section{Introduction}
\label{intro}

The interplay of quantum mechanical interference
with quantized energy exchange results in a quantum pump effect
which is investigated intensively both experimentally
\cite{SMCG99}$^{-}$\cite{VDM04}
and theoretically.
\cite{Thouless83}$^{-}$\cite{TB04}
This phenomenon being promising for manipulating and controlling
the passage of electrons through mesoscopic circuits
is of fundamental interest. Adiabatic driving involves
only low energy exchange and 
avoids excitations into inelastic channels which 
degrade the quantum properties of the system. 
In this work we investigate the 
magnetic symmetry properties of the dc-current of a quantum pump
which might operate in the presence of applied voltages and 
temperature gradients.

The experimentally measured adiabatically pumped dc current
\cite{SMCG99}
flowing through a chaotic cavity with periodically varying shape
is symmetric in magnetic field $H$.
That is in seeming contradiction with the theory
\cite{Brouwer98,ZSA99,SAA00,AAK00,VAA01}
predicting that the pumped current has no definite symmetry
under magnetic field reversal.
As a result it was conjectured
\cite{Brouwer01}$^{-}$\cite{MCM01}
that the current measured in Ref.~\onlinecite{SMCG99} is caused
by a classical rectification effect.
Indeed subsequent measurements \cite{DMH03} confirmed that
for slow {\it one-parameter} driving there is a symmetric in magnetic field
induced current whose origin is classical rectification.
Nevertheless one can not exclude the possibility that the current measured
in Ref.~\onlinecite{SMCG99} contains also the contribution coming from the
quantum pump effect.
To check it, perhaps, it is necessary to investigate the system
in a less symmetric setup, i.e., with reservoirs having different
electrochemical potentials or temperatures.
In the present paper we give a simple example when
the pumped current has or has not an odd in magnetic field contribution
depending on whether there is or there is no applied voltage.
Further experimental and theoretical efforts
to detect and distinguish the quantum pump effect
are highly desirable in view of a possible
application in quantum information processing devices. 
\cite{SamuelssonB04,BTT05}

The aim of the present paper is to explore in detail 
the symmetry properties of the
adiabatic current generated by the periodically driven mesoscopic conductor. 
To this end we represent the Floquet scattering matrix 
at low driving frequency $\omega$
as a sum of different terms with well defined symmetry properties 
(e.g., with respect to a magnetic field direction reversal).
One term reflects the symmetry of a stationary scattering process while the 
other term vanishing at $\omega\to 0$ has symmetry properties opposite 
to a stationary scattering process.
Based on such a representation we divide the dc current into parts
with well defined symmetry properties. 
That opens up additional possibilities for the experimental 
detection of the quantum pump effect.

In particular, in the two terminal case, 
we find a voltage dependent contribution to the pumped current which is odd in magnetic field. 
At small voltage this current is linear in $V$. Thus for small magnetic fields the 
dc-current has a component which is proportional to the product of frequency, magnetic field, and applied voltage.
For comparison we recall that 
in the stationary case, for a two-terminal conductor, the current linear in voltage (or, alternatively, the conductance) is an even function of a magnetic field.
\cite{Buttiker86,Buttiker92} 
A current that is odd in magnetic field appears only in the nonlinear 
voltage regime \cite{SB04,SZ04}
and is caused by electron-electron interactions.
In contrast, in the non-stationary case considered here even 
non-interacting electrons can show a response that is odd in magnetic field and
linear in applied voltage.

Recently the magnetic field symmetry of the dc current through 
an open quantum dot subject to a one-parameter potential oscillation has 
been investigated experimentally and theoretically as a function of frequency. 
\cite{VDM04}
In contrast, in the present paper we consider a two-parameter oscillation  
and investigate the magnetic field symmetry of the dc current in 
the presence of adiabatic parametric quantum pumping. 

The paper is organized as follows. 
In Sec.\ref{GA} we briefly consider the Floquet function approach
to scattering of electrons at a periodically driven mesoscopic conductor
and analyze the consequences of microscopic reversibility.
We introduce an exact representation for the scattering matrix 
at low driving frequency $\omega$. 
According to this representation the Floquet scattering matrix elements 
(up to linear in $\omega$ terms) are proportional to the elements of both 
the stationary scattering matrix $\hat S_{0}$ 
and a residual Floquet matrix $\hat A$ which exhibits symmetry
properties opposite to those of $\hat S_{0}$.
The symmetry properties of $\hat S_{0}$ are dictated by micro-reversibility,
and the residual Floquet matrix $\hat A$ reflects directly the breaking of these symmetries 
due to the driving of the sample.
Using such a representation we analyze the magnetic field symmetry of the dc 
current flowing through the adiabatically driven scatterer in Sec.\ref{MFS}.
We show that in the two terminal case
there is a dc current $I^{(od)}$ that is odd in magnetic field,
linear in $\omega$ and dependent on the applied voltage.
To calculate correctly $I^{(od)}$ it is necessary to find
the residual Floquet matrix $\hat A$.
Using several simple examples we outline the method for calculating
$\hat A$ in Sec.\ref{SE}.
We conclude  in Sec.\ref{C}.

\section{General approach}
\label{GA}

We use the scattering matrix approach
\cite{Buttiker90,Buttiker92,BTP94}
which views the mesoscopic sample as a scatterer which causes 
transmission and reflection of incident carriers. 
The scatterer is assumed to be coupled to $N_r$ 
reservoirs via single channel ballistic leads which we will number by
the Greek letters $\alpha, \beta$, etc.

We assume that in the stationary case 
electrons coming from the reservoirs and interacting with 
the scatterer are subject only to elastic scattering. 
Such (single particle) scattering can be described with the help 
of the scattering matrix $\hat S_{0}$. The index $0$ denotes
the stationary scattering matrix. In general $\hat S_{0}$
is a function of the electron energy $E$. 
This matrix collects all the quantum mechanical amplitudes for electrons
coming from some lead $\beta$ to be scattered into the same or any other lead
$\alpha$.
These amplitudes are normalized in such a way that their square define
the corresponding particle fluxes (currents).
If the electron velocities at a given energy
are the same in all the leads we can use these
amplitudes to relate the incident and out-going wave functions.
For instance, let
$\Psi^{(in)}_{0,\beta}(E,t) =
e^{-i\frac{E}{\hbar}t}\psi^{(in)}_{0,\beta}(E)$,
be the amplitude of a wave function describing 
electrons with energy $E$ incident in lead $\beta$.
Then the amplitude of the wave function of 
particles outgoing in lead $\alpha$,
$\Psi^{(out)}_{0,\alpha}(E,t) =
e^{-i\frac{E}{\hbar}t}\psi^{(out)}_{0,\alpha}(E)$,
is defined as follows:

\begin{equation}
\label{Eq1}
 \psi^{(out)}_{0,\alpha}(E) = \sum\limits_{\beta=1}^{N_r}
S_{0,\alpha\beta}(E) \psi^{(in)}_{0,\beta}(E).
\end{equation}

Current conservation implies that the scattering matrix
is a unitary matrix: \cite{LL3}
\begin{equation}
\label{Eq2}
 \hat S^{\dagger}_0 \hat S_0 =  \hat S_0 \hat S^{\dagger}_0 =  \hat I,
\end{equation}

\noindent where $\hat I$ is a unit matrix.
In fact, the knowledge of the matrix $\hat S_0(E)$ is equivalent to 
the knowledge of the solution for the stationary 
Schr\"odinger equation. 

For the dynamical problem with time-dependent scattering, 
scattering is characterized by 
the integral scattering operator which depends on two times \cite{VAA01}. 
One time-argument relates to the incoming
states and the second time-argument to the outgoing states. 
In this paper we are dealing
with a particular non stationary case, namely 
with a periodically driven scattering problem.
We assume that the scattering potential 
(hence the scattering properties of a sample) is varied in time periodically
with period ${\cal T} = 2\pi/\omega$. 
Then, according to the Floquet theorem 
(see, e.g., Refs.~\onlinecite{Shirley65}-\onlinecite{MBstrong02}),
the solution for the time-dependent Schr\"odinger equation 
can be represented in a relatively simple form
\begin{equation}
\label{Eq3}
\Psi(E,t) = e^{-i\frac{E}{\hbar}t} \sum\limits_{n=-\infty}^{\infty}
e^{-in\omega t} \psi_{}(E_n).
\end{equation}

\noindent 
Here $E$ is the Floquet energy;
$\psi(E_n)$ is a general solution of the stationary Schr\"odinger
equation corresponding to the energy $E_n = E + n\hbar\omega$.

Scattering on such an oscillatory scatterer can be
described via the Floquet scattering matrix. 
In this work we are concerned with the low-frequency 
properties of this dynamic problem and 
the relevant Floquet matrix $\hat S_{F}$ describes the
transitions between the propagating states only. \cite{MBstrong02} 
The elements $S_{F,\alpha\beta}(E_n,E)$ of this matrix are the quantum
mechanical amplitudes (normalized for current)
for an electron with energy $E$ to enter the
scatterer through lead $\beta$ and to leave the scatterer with
energy $E_n = E + n\hbar\omega$ through lead $\alpha$.  

In particular, if the reservoirs are stationary then 
the incoming wave function is 
$\Psi^{(in)}_{0,\beta}(E,t)$ and the wave function for particles outgoing 
to lead $\alpha$ is of the form Eq.(\ref{Eq3}) with 
\begin{equation}
\label{Eq4}
\psi_{\alpha}^{(out)}(E_n) = \sum\limits_{\beta=1}^{N_r}\sum\limits_{m}
\sqrt{\frac{k_m}{k_n}}S_{F,\alpha\beta}(E_{n},E_{m})
\psi^{(in)}_{0,\beta}(E_m)
\end{equation}

\noindent 
Here $k_{n}=\sqrt{2m_eE_{n}}/\hbar$ with $m_e$ being the electron mass.
Physically Eq.(\ref{Eq4}) means that an electron interacting with 
an oscillating  scatterer can gain or
lose one or several energy quanta $n\hbar\omega,~n = 0,\pm 1,\pm
2,\dots$, and thus an electron can change its energy by a discrete amount
$n\hbar\omega$.

Current conservation implies 
again that also the matrix $\hat S_{F}$ is unitary.
For the Floquet scattering matrix the analog of Eq.(\ref{Eq2}) 
reads as follows:
\begin{subequations}
\label{Eq5}
\begin{equation}
\label{Eq5A}
\sum\limits_{\alpha}\sum\limits_{n}
S^{*}_{F,\alpha\beta}(E_{n},E)S_{F,\alpha\gamma}(E_{n},E_{m})
= \delta_{m0}\delta_{\beta\gamma},
\end{equation}
\begin{equation}
\label{Eq5B}
\sum\limits_{\beta}\sum\limits_{n}
S^{*}_{F,\alpha\beta}(E,E_{n})S_{F,\gamma\beta}(E_{m},E_{n})
= \delta_{m0}\delta_{\alpha\gamma}.
\end{equation}
\end{subequations}
\noindent 
Here the summation over $n$ goes only over those $n$
which correspond to a positive $E_n = E + n\hbar\omega$.
In the low frequency limit we have $\hbar\omega\ll E$, 
and thus $n$ extends from $-\infty$ to $+\infty$.

To find the Floquet scattering matrix one needs to solve 
a fully time-dependent Schr\"odinger equation. Compared to the 
stationary problem, 
this is a more difficult  
and, generally, it can be done only numerically.  
On the other hand the representation Eq.(\ref{Eq3}) seems effectively to reduce 
the periodically driven case to the stationary one.
Therefore it is attractive to try to relate the Floquet scattering matrix 
$\hat S_{F}$ to the stationary scattering matrix $\hat S_{0}$.

\subsection{Adiabatic approximation}
\label{GAAA}

Let the stationary scattering matrix $\hat S_{0}(E,\{p\})$ 
depend on a set of parameters 
$p_i \in \{p\} , i = 1,2,\dots ,N_p$ (e.g., the sample's shape, the
strength of coupling to leads, the magnetic field, etc.). 
Varying these parameters one
can change the scattering properties of a sample. We take these
parameters to be periodic functions in time: 
$p_{i}(t) =p_{i}(t+{\cal T}) , \forall i$. 
Then the matrix $\hat S_{0}$ becomes time-dependent: 
$\hat S_{0}(E,t) = \hat S_{0}(E,\{p(t)\})$. 
In general the matrix $\hat S_{0}(t)$ does not describe the scattering of
electrons by a time-dependent scatterer: only the Floquet
scattering matrix $\hat S_{F}$ does.
Nevertheless in the low frequency limit, $\omega\to 0$, 
there exists a connection between these two matrices.
This connection becomes more evident 
if one represents the Floquet
scattering matrix elements as a series in powers of  $\omega$. 

\subsubsection{zeroth order approximation}
\label{GAAA0}

To zero-th order in the driving frequency 
the elements of the Floquet scattering matrix  $\hat S_{F}(E_n,E)$
can be approximated by the Fourier coefficients  
$\hat S_{0,n}$ of the stationary scattering matrix $\hat S_{0}$ 
as follows: \cite{MBstrong02}
\begin{subequations}
\label{Eq6}
\begin{equation} 
\label{Eq6A} 
\hat S_{F}(E_n,E) = \hat S_{0,n}(E) + O(\omega).  
\end{equation} 
\begin{equation} 
\label{Eq6B} 
\hat S_{F}(E,E_n) = \hat S_{0,-n}(E) + O(\omega).  
\end{equation} 
\end{subequations}

\noindent 
Here $O(\omega)$ denotes the rest which is at  
least first order in frequency $\omega$  and which is neglected
in the zero-th order adiabatic approximation. 
The Fourier transformation used reads as follows
\begin{subequations} 
\label{Eq7} 
\begin{equation} 
\label{Eq7A} 
\hat S_{0}(E,t) = \sum\limits_{n=-\infty}^{\infty}e^{-in\omega t} 
\hat S_{0,n}(E),  
\end{equation} 
\begin{equation} 
\label{Eq7B} 
\hat S_{0,n}(E) = \int\limits_{0}^{\cal T} \frac{dt}{{\cal T}}
e^{in\omega t}  \hat S_{0}(E,t). 
\end{equation} 
\end{subequations} 
\noindent
Before proceeding we check that this approximation is consistent
with the current conservation condition. 
Substituting Eq.(\ref{Eq6}) into Eq.(\ref{Eq5}) and performing
the inverse Fourier transformation we arrive at Eq.(\ref{Eq2}).

Equation (\ref{Eq6}) corresponds to the {\it frozen} scattering
matrix approximation. 
Within this approximation the stationary scattering matrix 
(with parameters dependent on time) completely characterizes 
the time-dependent scattering. 
This approximation is exact if the scattering matrix $\hat S_0$ is 
independent of the electron energy $E$ within the relevant energy interval.
\cite{MBstrong02}

\subsubsection{first order approximation}
\label{GAAA1}

To first order in the pump-frequency  $\omega$
we can represent the Floquet matrix with the help 
of the frozen scattering matrix, its energy derivatives and 
a matrix $\hat A$. 
In general the matrix $\hat A$ can not be expressed in terms of 
the stationary scattering matrix $\hat S_{0}$ 
and it has to be calculated (like $\hat S_{0}$ itself) 
in each particular case. The advantage of the representation 
which we introduce is that
the matrix $\hat A$ has a much smaller number of  elements 
than the Floquet scattering matrix.
The matrix $\hat A$ depends on only one energy, $E$, and therefore
it has $N_{r}\times N_{r}$ elements like the stationary scattering matrix 
$\hat S_{0}$. In contrast, 
the Floquet scattering matrix $\hat S_{F}$ depends on two energies, 
$E$ and $E_{n} = E + n\hbar\omega$, and therefore
has $\sim (2n_{max}+1)\times N_{r}\times N_{r}$ relevant elements. 
Here $n_{max}$ is the maximum number of energy quanta $\hbar\omega$
absorbed/emitted by an electron interacting with the scatterer 
which we should take into account to correctly describe the scattering process.
For small amplitude driving we have $n_{max}\approx 1$. In contrast, 
if the parameters vary with a large amplitude then  
$n_{max}\gg 1$. We represent the Floquet matrix in the form: \cite{MBac03}
\begin{subequations} 
\label{Eq8} 
\begin{equation} 
\label{Eq8A} 
\hat S_{F}(E_n,E) = \hat S_{0,n}(E)  
+ \frac{n\hbar\omega}{2}\frac{\partial\hat S_{0,n}(E)}{\partial E}  
+ \hbar\omega\hat A_{n}(E) + O(\omega^2),  
\end{equation} 
\begin{equation} 
\label{Eq8B} 
\hat S_{F}(E,E_n) = \hat S_{0,-n}(E)  
+ \frac{n\hbar\omega}{2}\frac{\partial\hat S_{0,-n}(E)}{\partial E}  
+ \hbar\omega\hat A_{-n}(E) + O(\omega^2).  
\end{equation} 
\end{subequations}
\noindent 
Note the right hand side (RHS) of Eq.(\ref{Eq8A}) is defined with respect to
the incoming energy of carriers, while in Eq.(\ref{Eq8B}) 
the RHS is expressed in terms of
the energy of outgoing particles. 
To first order in $\omega$, the case of interest here, 
these two representations are fully consistent. 
Going from one representation to the other, 
one needs to take into account that
the contribution from the first term on the RHS depends on the choice
of the reference energy. 
The second and the third terms being themselves proportional to $\omega$ 
do not depend on this choice.

In Eq.(\ref{Eq8}) we have introduced a new matrix 
$\hat A(E,t)$ with Fourier coefficients $\hat A_{n}(E)$. 
The current conservation condition, Eq.(\ref{Eq5}), leads to the 
following equation for the matrix $\hat A(E,t)$: \cite{MBac03}
\begin{subequations}
\label{Eq9}
\begin{equation}
\label{Eq9A}
\hbar\omega\left(\hat S^{\dagger}_{0}(E,t)\hat A(E,t) 
+ \hat A^{\dagger}(E,t)\hat S_{0}(E,t)\right)
= \frac{1}{2}{\cal P}\{\hat S^{\dagger}_{0};\hat S_{0} \},
\end{equation}
\begin{equation}
\label{Eq9B}
{\cal P}\{\hat S^{\dagger}_{0};\hat S_{0} \} =
i\hbar \left( \frac{\partial \hat S^{\dagger}_{0}}{\partial t}
\frac{\partial \hat S_{0}}{\partial E} -
\frac{\partial \hat S^{\dagger}_{0}}{\partial E}
\frac{\partial \hat S_{0}}{\partial t}
\right).
\end{equation}
\end{subequations}
Note the matrix ${\cal P}\{\hat S^{\dagger}_{0};\hat S_{0} \}$ is traceless. 
Another but equivalent representation can be obtained from Eq.(\ref{Eq9A})
multiplying both sides from the left by $\hat S_{0}$ and from the right by 
$\hat S_{0}^{\dagger}$, and by taking into account that because of the unitarity
condition, Eq.(\ref{Eq2}), we have 
$S_{0}d[S^{\dagger}_{0}] = - d[S_{0}]S^{\dagger}_{0}$.

We remark that Eq.(\ref{Eq9}) tells us that the expansion in powers of
$\omega$ is, in fact, an expansion in powers of $\hbar\omega/\delta E$,
where $\delta E$ is the energy scale 
over which the scattering matrix $\hat S_{0}(E)$ changes significantly.
Therefore, the frequency $\omega$ can be considered as slow and the
expansion Eq.(\ref{Eq8}) can be relevant if
 \begin{equation}
\label{Eq10}
 \hbar\omega \ll \delta E.
\end{equation}
Consequently, to characterize scattering with an accuracy of order $\omega$ 
one needs to determine the matrix $\hat A$. 
Equation (\ref{Eq9}) defines only the anti commutator of two matrices,
$\hat S_{0}$ and $\hat A$, and it is insufficient to determine
the matrix $\hat A$.

By analogy with Eq.(\ref{Eq6}) we can express the Floquet scattering matrix
elements up to first order in driving frequency 
in terms of the Fourier coefficients of some effective matrix.
We introduce two matrices $\hat S_{in}$ and $\hat S_{out}$ defined 
with respect to incoming and outgoing energies, respectively:
\begin{subequations}
\label{Eq11}
\begin{equation} 
\label{Eq11A}
\hat S_{in}(E,t) = \hat S_{0}(E,t) + \frac{i\hbar}{2}
\frac{\partial^2\hat S_{0}}{\partial t\partial E} + \hbar\omega\hat A(E,t).  
\end{equation} 
\begin{equation} 
\label{Eq11B}
\hat S_{out}(E,t) = \hat S_{0}(E,t) - \frac{i\hbar}{2}
\frac{\partial^2\hat S_{0}}{\partial t\partial E} + \hbar\omega\hat A(E,t).  
\end{equation} 
\end{subequations}
Performing the Fourier transformation of Eqs.(\ref{Eq11})
and comparing the result with Eqs.(\ref{Eq8}) we find:
\begin{subequations}
\label{Eq12}
\begin{equation} 
\label{Eq12A} 
\hat S_{F}(E_n,E) = \hat S_{in,n}(E) + O(\omega^2).  
\end{equation} 
\begin{equation} 
\label{Eq12B} 
\hat S_{F}(E,E_n) = \hat S_{out,-n}(E) + O(\omega^2).  
\end{equation} 
\end{subequations}
We emphasize that the matrices $\hat S_{in}(t)$ and $\hat S_{out}(t)$ 
are not {\it scattering} matrices because they are not unitary:
Their Fourier coefficients just define the corresponding matrix elements
of the Floquet scattering matrix according to Eq.(\ref{Eq12}).
Nevertheless these matrices conserve the current "on average", 
i.e. after integrating over the time period ${\cal T}$: 
\begin{subequations}
\label{Eq13}
\begin{equation}
\label{Eq13A}
\int\limits_{0}^{\cal T} \frac{dt}{{\cal T}} 
\hat S_{in}^{\dagger}(E,t) \hat S_{in}(E,t) = \hat I + O(\omega^2).
\end{equation}
\begin{equation}
\label{Eq13B}
\int\limits_{0}^{\cal T} \frac{dt}{{\cal T}} 
\hat S_{out}^{\dagger}(E,t) \hat S_{out}(E,t) = \hat I + O(\omega^2).
\end{equation}
\end{subequations}
Now we use Eq.(\ref{Eq9}) to analyze the general properties of 
the matrix $\hat A$ 
which are due to the micro reversibility of the Schr\"odinger equation
with a periodically oscillating potential.

\subsection{Micro-reversibility and magnetic field symmetry 
                   of the Floquet scattering matrix}
\label{GAMM}

We start with the stationary case when the single particle Hamiltonian 
(and correspondingly the scattering matrix) is independent of time
and recall some properties of the stationary scattering matrix. 
\cite{LL3,Buttiker92}

The micro-reversibility of the equation of motion (i.e., the Schr\"odinger equation)
puts some constraints onto the scattering matrix. 
To make the notation more convenient let us arrange the incoming/outgoing 
wave functions at all the leads into the vector column
\begin{equation}
\label{Eq14}
\hat\psi = \left(
\begin{array}{c}
\psi_{1} \\
\psi_{2} \\
\vdots \\
\psi_{N_r}
\end{array}
\right).
\end{equation}
Then Eq.(\ref{Eq1}) can be written in the compact form:
\begin{equation}
\label{Eq15}
\hat\psi^{(out)} = \hat S_{0}\hat\psi^{(in)}.
\end{equation}
The micro-reversibility condition 
(i.e., the invariance with respect to the time inversion)
for the spin less case under consideration 
leaves the solution of the scattering problem invariant
under the simultaneous inversion of the direction of movement,
the inversion of a possibly present magnetic field $H$,
and the replacement $\Psi \to  \Psi^{*}$.
Therefore, the evolution of the two wave functions, namely 
$\Psi(E,H,t)$ and $\Psi^{*}(E,-H,-t)$, 
is exactly the same and is described by the same scattering matrix $\hat S_0$. 
Taking into account 
that the inversion of the direction of movement turns the outgoing
waves to incoming ones and vice versa we can write 
the following equations for the starting solution and its transform:
\begin{subequations}
\label{Eq16}
\begin{equation}
\label{Eq16A}
\hat\psi^{(out)}(E,H) = \hat S_{0}(E,H)\hat\psi^{(in)}(E,H), 
\end{equation}
\begin{equation}
\label{Eq16B}
\left(\hat\psi^{(in)}(E,-H)\right)^{*} = 
\hat S_{0}(E,H)\left(\hat\psi^{(out)}(E,-H)\right)^{*}. 
\end{equation}
\end{subequations}
From the unitarity condition Eq.(\ref{Eq2}) 
it follows that $\hat S_{0}^{-1} = \hat S_{0}^{\dagger}$.
Therefore we can rewrite Eq.(\ref{Eq16A}) as follows: 
$\hat\psi^{(in)}(E,H) = \hat S_{0}^{\dagger}(E,H)\hat\psi^{(out)}(E,H)$.
Comparing the last with Eq.(\ref{Eq16B}) we arrive at the required condition:
\cite{Buttiker92}
\begin{equation}
\label{Eq17}
\hat S_{0}(-H) = \hat S_{0}^{T}(H),
\end{equation}

\noindent
where the upper index "$T$" denotes transposition.

Next we consider a periodically driven scattering problem. 
As we saw micro-reversibility requires
the  scattering matrix to be symmetric with respect to the interchange
of incoming and outgoing channels. 
For the Floquet scattering matrix these channels are characterized
by both the lead index and the number $n$ showing how many
energy quanta $\hbar\omega$ an electron absorbs/emits during 
the scattering process. 
In addition, to get the required symmetry condition, we have to 
take into account that
the parameters $p_i$ of the Hamiltonian depend on time.
We suppose they change periodically in time with the same frequency 
$\omega$, and with possible relative phase shifts $\varphi_i$:
\begin{equation}
\label{Eq18}
 p_i(t) = p_{i,0} + p_{i,1}\cos(\omega t + \varphi_i).
\end{equation}
In such a case time reversal implies the inversion of the sign of all the
phase shifts $\varphi_i$. 
Therefore, the Floquet scattering matrix elements
are subject to the following fundamental symmetry:
\begin{subequations}
\label{Eq19}
\begin{equation}
\label{Eq19A}
S_{F,\alpha\beta}(E,E_{n}; H, \varphi) = 
S_{F,\beta\alpha}(E_{n},E; -H, -\varphi), 
\end{equation}
{\rm or in a matrix form}
\begin{equation}
\label{Eq19B}
\hat S_{F}(E; -H, -\varphi) = \hat S_{F}^{T}(E; H, \varphi).
\end{equation}
\end{subequations}

\noindent 
Here $E$ is the Floquet energy [see, Eq.(\ref{Eq2})];
$\varphi$ denotes the set of all the $\varphi_i$.

Next we derive the  symmetry conditions for the matrix $\hat A$
entering  Eq.(\ref{Eq8}). 
Our definition of the phases $\varphi_i$ 
[see, Eq.(\ref{Eq18})] implies that 
the frozen scattering matrix $\hat S_{0}(E,t)$ 
[i.e., the stationary scattering matrix with parameters dependent on time
$\hat S_{0}(E,t) = \hat S_{0}(E,p_i(t))$ ] possesses the following symmetry
\begin{equation}
\label{Eq20}
\hat S_{0}(E, -t;  H, -\varphi) = \hat S_{0}(E, t; H, \varphi).
\end{equation}
Then equation (\ref{Eq9}) gives us:
\begin{equation}
\label{Eq21}
\hat A(E, -t; H, -\varphi) = - \hat A(E, t; H, \varphi).
\end{equation}
Correspondingly, for the Fourier coefficients, we have the following:
\begin{subequations}
\label{Eq22}
\begin{equation}
\label{Eq22A}
\hat S_{0,n}(E; H, -\varphi) = \hat S_{0,-n}(E; H, \varphi),
\end{equation}
\begin{equation}
\label{Eq22B}
\hat A_{n}(E; H, -\varphi) = - \hat A_{-n}(E; H, \varphi).
\end{equation}
\end{subequations}
Substituting the equations given above into 
the adiabatic expansion, Eqs.(\ref{Eq8}),
and taking into account the micro-reversibility condition, Eq.(\ref{Eq19}),
we find the required symmetry condition for the matrix 
$\hat A(t)$:
\begin{equation}
\label{Eq23}
\hat A(-H) = - \hat A^{T}(H).
\end{equation}
In particular, in the absence of magnetic fields, $H=0$, 
the diagonal elements of $\hat A$ vanish. 
That was previously shown in Ref.~\onlinecite{MBac03}. 
Alternatively equation (\ref{Eq23}) can be obtained directly from
Eq.(\ref{Eq9}) exploiting the symmetry condition Eq.(\ref{Eq17})
and the unitarity of the frozen scattering matrix $\hat S_{0}(E,t)$.

The symmetry properties of the residual Floquet matrix $\hat A$ are completely different 
from that of the stationary scattering matrix $\hat S_{0}$.
The residual Floquet matrix $\hat A$ reflects directly the most important differences 
between an adiabatic scattering process at a periodically evolving scatterer
and a strictly stationary scattering process.

\section{Magnetic field symmetry of the dc current
flowing through the slowly driven scatterer}
\label{MFS}

Now we use the results of the previous section to analyze 
the dc current through the mesoscopic sample with 
periodically varying parameters.
We will consider two mechanisms which can give rise to such a current. 
The first mechanism is a quantum pump effect 
consisting in rectifying of time-dependent currents generated by
the non stationary scatterer. \cite{MBac03}
Second we permit a constant in time difference of electrochemical 
potentials/temperatures between the different reservoirs.
The last is important, because the widely investigated situation 
with reservoirs being at the same electrochemical potential actually 
hides some physics underlying the quantum pump effect.

The dc current $I_{\alpha}$ flowing from the scatterer to the reservoir
in the lead $\alpha$ can be calculated as follows: \cite{MBstrong02}
\begin{equation}
\label{Eq24}
I_{\alpha} = \frac{e}{h} \int\limits_{0}^{\infty} dE 
\left\{
\sum\limits_{\beta=1}^{N_r} \sum\limits_{n}
\left| S_{F,\alpha\beta}(E_n,E) \right|^2 f_{0,\beta}(E) - f_{0,\alpha}(E)
\right\}.
\end{equation}
\noindent
Here $f_{0,\alpha}$ is the electron distribution function for 
the reservoir $\alpha$. 
We assume that the reservoirs are in a stationary equilibrium state
with possibly different electrochemical potentials $\mu_{\alpha}$ 
and temperatures $T_{\alpha}$. 
Then $f_{0,\alpha}$ is the Fermi distribution function
\begin{equation}
\label{Eq25}
f_{0,\alpha}(E) = \frac{1}{1+ e^{ \frac{E - \mu_{\alpha}}{k_BT_{\alpha}}}},
\end{equation}
\noindent 
with $k_B$ being the Boltzmann constant.
Substituting the adiabatic expansion Eq.(\ref{Eq8}) into Eq.(\ref{Eq24})
and performing the inverse Fourier transformation
we find the current up to linear in $\omega$ terms as follows: \cite{MBac03}
\begin{equation}
\label{Eq26}
\begin{array}{c}
I_{\alpha} =   \int\limits_{0}^{\infty} dE  
\int\limits_{0}^{\cal T} \frac{dt}{{\cal T}}
\sum\limits_{\beta}  
\left\{   
f_{0,\beta}(E) \frac{dI_{\alpha\beta}(E,t)}{dE}
 \right. \\
\ \\
\left.
+  \frac{e}{h} \big|S_{0,\alpha\beta}(E,t)\big|^2  
\big[ f_{0,\beta}(E) - f_{0,\alpha}(E) \big] 
\right\},  
\end{array}
\end{equation}

\noindent
where $dI_{\alpha\beta}/dE$ is a spectral current driven by the non stationary 
scatterer from lead $\beta$ into lead $\alpha$: 
\begin{equation} 
\label{Eq27} 
\frac{dI_{\alpha\beta}}{dE} = \frac{e}{h}\Big(
2\hbar\omega Re[S^{*}_{0,\alpha\beta}A_{\alpha\beta}]
+\frac{1}{2}{\cal P}\{S_{0,\alpha\beta}; S^{*}_{0,\alpha\beta}\} \Big). 
\end{equation}

\noindent Here $Re[X]$ is the real part of $X$; 
the function ${\cal P}\{X;Y\}$ is defined in Eq.(\ref{Eq9B}).
The spectral currents $dI_{\alpha\beta}/dE$ are subject to the following 
conservation law: \cite{MBac03}
\begin{equation}
\label{Eq28}
\sum\limits_{\alpha=1}^{N_r} \frac{dI_{\alpha\beta}(E,t)}{dE} = 0,
\end{equation}
Using Eq. (\ref{Eq28}) and the unitarity of the frozen scattering matrix,
$\sum_{\alpha} |S_{0,\alpha\beta}|^2 = \sum_{\beta}  |S_{0,\alpha\beta}|^2 =1$,
one can easily check that 
the current $I_{\alpha}$ is conserved: $\sum_{\alpha}I_{\alpha} = 0$.
Further, using the symmetry conditions,
Eqs.(\ref{Eq17}) and (\ref{Eq23}),
and rearranging the terms in Eq.(\ref{Eq26}) 
we divide the current into the even, 
$I_{\alpha}^{(ev)}(H) = I_{\alpha}^{(ev)}(-H)$,
and odd,
$I_{\alpha}^{(od)}(H) = -I_{\alpha}^{(od)}(-H)$,
in magnetic field parts:
\begin{subequations}
\label{Eq29}
\begin{equation}
\label{Eq29A} 
\begin{array}{c}
I_{\alpha}^{(ev)}(H) =   \frac{e}{h} \int\limits_{0}^{\infty} dE  
\int\limits_{0}^{\cal T} \frac{dt}{{\cal T}}\sum\limits_{\beta} 
\bigg\{  [f_{0,\beta} - f_{0.\alpha}] \\
\times
\Big( \frac{|S_{0,\alpha\beta}|^2  + |S_{0,\beta\alpha}|^2 }{2} 
+ \hbar\omega Re[S^{*}_{0,\alpha\beta}A_{\alpha\beta}
- S^{*}_{0,\beta\alpha}A_{\beta\alpha}] \Big) \\
+ [f_{0,\beta} + f_{0,\alpha}]
\frac{{\cal P}\{S_{0,\alpha\beta}; S^{*}_{0,\alpha\beta}\} 
+ {\cal P}\{S_{0,\beta\alpha}; S^{*}_{0,\beta\alpha}\}}{4}\bigg\},
\end{array}
\end{equation}
\begin{equation}
\label{Eq29B}
\begin{array}{c}
I_{\alpha}^{(od)}(H) =   \frac{e}{h} \int\limits_{0}^{\infty} dE  
\int\limits_{0}^{\cal T} \frac{dt}{{\cal T}}\sum\limits_{\beta} 
\bigg\{  [f_{0,\beta} - f_{0.\alpha}] \\
\times
\Big( \frac{|S_{0,\alpha\beta}|^2  - |S_{0,\beta\alpha}|^2 }{2} 
+ \frac{{\cal P}\{S_{0,\alpha\beta}; S^{*}_{0,\alpha\beta}\} 
- {\cal P}\{S_{0,\beta\alpha}; S^{*}_{0,\beta\alpha}\} }{4}
\Big) \\
+ [f_{0,\beta} + f_{0,\alpha}]\hbar\omega 
Re[S^{*}_{0,\alpha\beta}A_{\alpha\beta}
+ S^{*}_{0,\beta\alpha}A_{\beta\alpha}]\bigg\}. 
\end{array}
\end{equation}
\end{subequations}
To show that both these currents are separately conserved,
i.e., that 
$\sum\limits_{\alpha}I_{\alpha}^{(ev)} = 0$
and $\sum\limits_{\alpha}I_{\alpha}^{(od)} = 0$,
one can use 
the relations \cite{MBac03}
\begin{subequations}
\label{Eq30} 
\begin{equation} 
\label{Eq30A} 
4\hbar\omega \sum\limits_{\alpha=1}^{N_r}
Re[S^{*}_{0,\alpha\beta}A_{\alpha\beta}] =  
{\cal P}\{\hat S_{0}^{\dagger};\hat S_{0} \}_{\beta\beta}, 
\end{equation} 
\begin{equation} 
\label{Eq30B} 
4\hbar\omega \sum\limits_{\beta=1}^{N_r}
Re[S^{*}_{0,\alpha\beta}A_{\alpha\beta}] =  
{\cal P}\{\hat S_{0};\hat S_{0}^{\dagger} \}_{\alpha\alpha}. 
\end{equation} 
\end{subequations}
which follow from Eq.(\ref{Eq9A}). 

In a general multi-terminal situation, 
i.e., if  not all the reservoirs are at the same potential (temperature),
the main contributions to both the even $I_{\alpha}^{(ev)}$ and the odd
$I_{\alpha}^{(od)}$ currents are proportional to the conductances 
$\frac{e^2}{h}|S_{0,\alpha\beta}|^2$ averaged over time. 
The non-stationarity results only in small corrections.
However in the two terminal case the odd in magnetic field dc current 
$I^{(od)}$
has no contribution coming from the conductances.
The current $I^{(od)}$ is linear in $\omega$ and 
it is entirely due to the non-adiabaticity of the pump scattering processes.

\subsection{Two terminal many channel scatterer}
\label{MFS2}

To show this let us consider the scatterer connected to only two reservoirs
via, possibly many channel, ballistic leads.
We will mark the quantities related to the left and to the right reservoirs 
via the lower indices "$L$" and "$R$", respectively.
Let the left lead have $N_{L}$ channels, and
the right lead have $N_{R}$ channels: $N_L + N_R = N_r$. 
We define the currents flowing to the left $I_{L}$
and to the right $I_{R} = - I_{L}$,
and the distribution functions 
for the left $f_{0,L}$ and for the right $f_{0,R}$ reservoirs
as follows:
$$
\begin{array}{l}
I_{L} = \sum\limits_{\alpha=1}^{N_L} I_{\alpha}, 
\quad \quad \quad
f_{0,\alpha} = f_{0,L}, \quad 1 \leq \alpha \leq N_L,  \\
\ \\
I_{R} = \sum\limits_{\alpha=N_L+1}^{N_r} I_{\alpha}, 
\quad
f_{0,\alpha} = f_{0,R}, \quad N_L+1 \leq \alpha \leq N_r.
\end{array}
$$
By analogy we redefine the quantities dependent on two indices. 
For example, the reflection to the left $R_{LL}$ and 
the spectral current $dI_{RL}/dE$ driven from the left to the right 
are defined as follows:
$$
\begin{array}{l}
R_{LL} = \sum\limits_{\alpha=1}^{N_L} \sum\limits_{\beta=1}^{N_L} 
|S_{0,\alpha\beta}|^2 ,\\
\ \\
\frac{dI_{RL}}{dE} =
\sum\limits_{\alpha=N_L+1}^{N_r} \sum\limits_{\beta=1}^{N_L} 
\frac{dI_{\alpha\beta}}{dE}.
\end{array}
$$
Note that the two terminal transmission is symmetric in
reservoirs indices, $T_{LR} = T_{RL}$, and it is even in magnetic field.
That can be easily seen from their definition, similar 
to the one given above for $R_{LL}$,  and from the unitarity
of the scattering matrix $\hat S_0$, Eq.(\ref{Eq2}).
In addition from Eq.(\ref{Eq28}) we get:
$dI_{L\xi}/dE + dI_{R\xi}/dE = 0$, for $\xi = L, R$. 
Using the identity: \cite{MBac03}
$$
\frac{dI_{LL}}{dE} + \frac{dI_{LR}}{dE} \equiv 
\frac{dI_{L}}{dE} =  \sum\limits_{\alpha=1}^{N_L} i\frac{e}{2\pi}  
\left(  \frac{\partial\hat S_{0}}{\partial t}
\frac{\partial\hat S_{0}^{\dagger}}{\partial E} -  
\frac{\partial\hat S_{0}}{\partial E}
\frac{\partial\hat S_{0}^{\dagger}}{\partial t} 
\right)_{\alpha\alpha}, 
$$
performing necessary summations in Eqs.(\ref{Eq29}),
and integrating by parts over time and over energy,
we get:
\begin{subequations}
\label{Eq31}
\begin{equation}
\label{Eq31A}
\begin{array}{c}
I_{L}^{(ev)} = \frac{e}{h} \int\limits_{0}^{\infty} dE  
\int\limits_{0}^{\cal T} \frac{dt}{{\cal T}} \bigg\{  [f_{0,R}-f_{0,L}]  \\
\times \Big( T_{LR} 
+  \sum\limits_{\alpha=1}^{N_L}\sum\limits_{\beta=N_L+1}^{N_r}
\frac{ {\cal P}\{S_{0,\alpha\beta}; S^{*}_{0,\alpha\beta}\} 
+ {\cal P}\{S_{0,\beta\alpha}; S^{*}_{0,\beta\alpha}\} }{4} \Big)  \\
+ \left(-\frac{\partial}{\partial E} [f_{0,R}+f_{0,L}]\right)  
\frac{i\hbar}{4} \sum\limits_{\alpha=1}^{N_L}\left(
\frac{\partial \hat S_{0}}{\partial t}\hat S_{0}^{\dagger} +
\hat S_{0}^{\dagger} \frac{\partial \hat S_{0}}{\partial t} \right)_{\alpha\alpha}
 \bigg\},
\end{array}
\end{equation}
\begin{equation}
\label{Eq31B}
\begin{array}{c}
I_{L}^{(od)} = I^{(od,ev)}_{L}+I^{(od,od)}_{L}, \\
I^{(od,ev)}_{L} = \frac{ei}{8\pi} \int\limits_{0}^{\infty} dE  
\int\limits_{0}^{\cal T} \frac{dt}{{\cal T}} 
\Big\{\left(-\frac{\partial}{\partial E} [f_{0,R}+f_{0,L}]\right)  \\
\times \sum\limits_{\alpha=1}^{N_L}
\left(\frac{\partial \hat S_{0}}{\partial t}\hat S_{0}^{\dagger} -
\hat S_{0}^{\dagger} \frac{\partial \hat S_{0}}{\partial t} \right)_{\alpha\alpha}
\Big\}, \\
I^{(od,od)}_{L}= \frac{e\omega }{2\pi} \int\limits_{0}^{\infty} dE  
\int\limits_{0}^{\cal T} \frac{dt}{{\cal T}} 
\Big\{[f_{0,R} - f_{0,L}]  \\
\times \sum\limits_{\alpha=1}^{N_L}\sum\limits_{\beta=N_L+1}^{N_r}
Re[S^{*}_{0,\alpha\beta}A_{\alpha\beta}
+ S^{*}_{0,\beta\alpha}A_{\beta\alpha}]\Big\}.
\end{array}
\end{equation}
\end{subequations}
For low driving frequencies, $\omega \to 0$,
we see that in the two terminal case the part of the dc-current that
is odd in magnetic field, 
$I^{(od)}(H) = - I^{(od)}(-H)$, is linear 
in $\omega$ irrespective of whether the reservoirs are at the same conditions 
($f_{0,L} = f_{0,R}$) or not ($f_{0,L} \neq f_{0,R}$).

Let us introduce the voltage $V$ and the temperature difference $\Delta T$ 
applied to the system, both constant in time:
\begin{equation}
\label{Eq32}
\begin{array}{c}
\mu_{R} = \mu_0 + \frac{eV}{2}, \quad \mu_{L} = \mu_0 - \frac{eV}{2},  \\
\ \\
T_{R} = T_0 + \frac{\Delta T}{2}, \quad  T_{L} = T_0 - \frac{\Delta T}{2},
\end{array}
\end{equation}

\noindent
and analyze the current $I^{(od)}$ in more detail.
According to Eq.(\ref{Eq31B}) this current consists of two parts,
$I^{(od)}_{L} = I^{(od,ev)}_{L} + I^{(od,od)}_{L}$.
The first one 
$$
I^{(od,ev)}_{L}(V,\Delta T) = I^{(od,ev)}_{L}(-V,-\Delta T), 
$$
is even in both $V$ and $\Delta T$ 
and it survives even at $f_{0,L} = f_{0,R}$. 
This contribution is due to conventional quantum pump effect.
\cite{Brouwer98}
In contrast, the second part
$$
I^{(od,od)}_{L}(V,\Delta T) = -I^{(od,od)}_{L}(-V,-\Delta T),
$$
is odd in both $V$ and $\Delta T$. 
Both contributions, $I^{(od,ev)}$ and $I^{(od,od)}$, have the same origin: 
They are rectified ac currents with spectral density 
$dI_{\alpha\beta}/dE$, Eq.(\ref{Eq27}), 
pushed by the pump from one reservoir to another.
The part $I^{(od,ev)}$ emphasizes the contribution arising if 
there are incoming electrons from both leads, $\alpha$ and $\beta$.
While the part $I^{(od,od)}$ is entirely due to an asymmetry
in electron flows incident from the leads. 
This asymmetry, due to the difference between the reservoir's 
distribution functions $f_{0,\alpha}$ and $f_{0,\beta}$, vanishes in the 
absence of an applied voltage, $V=0$, and in the absence of a 
temperature difference, $\Delta T = 0$.

For further reference we now give the equations (\ref{Eq31})
for the particular case of a scatterer connected to one-channel leads.

\subsubsection{Two terminal single channel scatterer}
\label{MFS2S}

For single channel leads, $N_L = N_R = 1$, the stationary scattering matrix $\hat S_{0}$ is a unitary $2\times 2$  matrix:
\begin{equation} 
\label{Eq33} 
\hat S_{0} = e^{i\gamma}\left( 
  \begin{array}{cc} 
       \sqrt{R}e^{-i\theta}   & i\sqrt{T} e^{-i\phi}           \\ 
       i\sqrt{T}e^{i\phi}      & \sqrt{R}e^{i\theta} \\ 
  \end{array} 
\right). 
\end{equation} 
 
\noindent 
Here $R$ and $T$ are the reflection and the transmission 
probability, respectively ($R+T=1$).  
The phase $\theta$ characterizes the asymmetry between  
the reflection to the left and to the right.  
The phase $\gamma$ relates to the charge on the scatterer.
The phase $\phi$ characterizes the asymmetry between 
the transmission through the scatterer from the left to the right and back
and it relates to the magnetic flux on the scatterer.

We assume that all these quantities are functions of
the electron energy $E$, the magnetic field $H$, and the external parameters  
$p_i(t)$ varying with frequency $\omega$.  
From Eq. (\ref{Eq17}) it follows that 
$R, T, \gamma$, and $\theta$ are even functions of the magnetic field $H$,
while $\phi$ is an odd function of $H$.

Using the scattering matrix, Eq.(\ref{Eq33}),
we rewrite the dc current 
$I_L =  - I_R$, Eq.(\ref{Eq31}), as follows:

\begin{subequations}
\label{Eq34}
\begin{equation}
\label{Eq34A}
I_L = I_{L}^{(ev,ev)} + I_{L}^{(ev,od)} 
+ I_{L}^{(od,ev)} + I_{L}^{(od,od)}, 
\end{equation}
\begin{equation}
\label{Eq34B}
I_{L}^{(ev,ev)} = \frac{e}{4\pi} \int\limits_{0}^{\infty} dE  
\int\limits_{0}^{\cal T} \frac{dt}{{\cal T}} 
\left(-\frac{\partial}{\partial E} [f_{0,R}+f_{0,L}]\right)  
 R \frac{\partial\theta}{\partial t},
\end{equation}
\begin{equation}
\label{Eq34C}
I_{L}^{(ev,od)} =  \frac{e}{h} \int\limits_{0}^{\infty} dE  
\int\limits_{0}^{\cal T} \frac{dt}{{\cal T}} [f_{0,R}-f_{0,L}]
\bigg(T - \frac{\hbar}{2}\frac{\partial}{\partial E}
\Big[ T \frac{\partial\gamma}{\partial t} \Big]  \bigg),
\end{equation}
\begin{equation}
\label{Eq34D}
I_{L}^{(od,ev)}  = \frac{e}{4\pi} \int\limits_{0}^{\infty} dE  
\int\limits_{0}^{\cal T} \frac{dt}{{\cal T}} 
\left(-\frac{\partial}{\partial E} [f_{0,R}+f_{0,L}]\right)  
T \frac{\partial\phi}{\partial t},
\end{equation}
\begin{equation}
\label{Eq34E}
I^{(od,od)}_{L}= \frac{e\omega}{2\pi} \int\limits_{0}^{\infty} dE  
\int\limits_{0}^{\cal T} \frac{dt}{{\cal T}} 
[f_{0,R} - f_{0,L}] Re[S^{*}_{0,LR}A_{LR} + S^{*}_{0,RL}A_{RL}]. 
\end{equation}
\end{subequations}

\noindent
Here the first upper index, $ev/od$, relates to the magnetic field
symmetry of the current,
while the second upper index relates to the symmetry with respect
to the applied voltage (temperature) difference.

The currents $I_{L}^{(ev,ev)}$ and $I_{L}^{(od,ev)}$
are conventional pumped currents 
(with reservoirs being at the same conditions).
They depend on the asymmetry of the stationary scattering matrix:
The phases $\theta$ and $\phi$ describe the asymmetry in the reflection from 
and in the transmission through the scatterer, respectively. 

The remaining two contributions,
$I_{L}^{(ev,od)}$ and $I_{L}^{(od,od)}$,
are present if the electron flows incoming from the reservoirs are different.
The even in magnetic field current $I_{L}^{(ev,od)}$ 
exists already in the stationary case.
The variation of the scattering parameters results in averaging over 
the time period and in the correction to the frozen conductance. \cite{EWAL02} 
In contrast the odd in magnetic field current $I_{L}^{(od,od)}$ 
exists only in the non-stationary regime. 

The contributions 
$I_{L}^{(ev,ev)}$, $I_{L}^{(od,ev)}$, $I_{L}^{(od,od)}$,
and the part (proportional to $\partial\gamma/\partial t$)
of $I_{L}^{(ev,od)}$ all are due to the quantum rectification of
ac currents, Eq.(\ref{Eq27}), generated by the oscillating scatterer.
This mechanism does work (i.e., a dc current exists) if the time reversal 
invariance is broken in the system by the varying parameters $p_i$
and hence the integral over the time period does not vanish.

The last statement is not evident for the current  $I_{L}^{(od,od)}$.
To make it clear we note the following: 
Generally the non-adiabatic corrections to the frozen solution of the Schr\"odinger
equation (and to the frozen scattering matrix) are proportional to the 
time derivative. 
The matrix $\hat A$ being a part of these corrections should be proportional 
to $\partial/\partial t$ as well.
Therefore, the conditions for $I_{L}^{(od,od)},$ Eq.(\ref{Eq34E}),
to be non-vanishing, (which requires that the integral 
over the time period is nonzero),
are generally the same as that for, e.g. $I_{L}^{(ev,ev)}$, Eq.(\ref{Eq34B}).

However, we emphasize that the necessary conditions to get 
a pumped current include both the time reversal symmetry breaking
and the presence of a spatial asymmetry. 
Strictly speaking these conditions are not identical for all the parts,
Eqs.(\ref{Eq34B}) - (\ref{Eq34E}).
In particular, the applied voltage $V$ makes the whole system 
(the sample plus reservoirs) to be spatially non-symmetric,
in the sense that the direction from the left to the right 
is not identical to the opposite direction.
Therefore, e.g., the current $I_{L}^{(od,od)}$ is less sensitive
to the spatial asymmetry of a scatterer than, e.g., the current 
$I_{L}^{(od,ev)}$ is.

In the next section we calculate the residual Floquet 
matrix $\hat A$ for several simple examples and 
illustrate the validity of the general statements made here.
In particular we consider
a one-dimensional loop with two leads and with enclosed magnetic flux.
This example shows that the current $I_{L}^{(od,od)}$
exists (at nonzero magnetic flux) 
already if only the time reversal invariance is broken
(i.e., if two parameters oscillate with a phase lag $\Delta\varphi\neq 0$).
In contrast, a non-zero current $I_{L}^{(od,ev)}$
requires (in addition to $\Delta\varphi\neq 0$) an asymmetry 
in coupling to the leads.  
If such a coupling is symmetric then 
the current $I_{L}^{(od,ev)}$ is identically zero no matter 
how much the magnetic flux through the ring is and whether 
$\Delta\varphi$ is zero or not.

\section{The Floquet scattering
matrix for low driving frequencies: Simple examples}
\label{SE}

In this section we illustrate 
how one can calculate the linear in $\omega$ corrections to the frozen scattering
matrix in the same fashion as the stationary scattering matrix $\hat S_{0}$. 

According to Eq.(\ref{Eq12}), at $\omega\to 0$
the Floquet scattering matrix elements are the Fourier coefficients of
some matrices $\hat S_{in}$/$\hat S_{out}$, Eq.(\ref{Eq11}) 
which depend on the stationary scattering matrix $\hat S_{0}$ 
and on the matrix $\hat A$.
The matrix $\hat S_{in}$/$\hat S_{out}$ does not possess a definite symmetry,
i.e., with respect to a time and/or a magnetic field direction reversal. 
While the stationary scattering matrix $\hat S_{0}$ and the matrix $\hat A$ do.
The last circumstance is the motivation why we expressed the current 
$I_{\alpha}$ in terms of $\hat S_{0}$ and $\hat A$ instead of 
$\hat S_{in}$/$\hat S_{out}$.
On the other hand for the calculation of the Floquet scattering matrix elements
it is more convenient to work in terms of the matrix 
$\hat S_{in}$ or $\hat S_{out}$.

\subsection{Single $\delta$-function barrier}
\label{SED}

In the first example we consider the Floquet scattering matrix in the limit $\omega\to 0$ 
of an oscillating  point-like scatterer coupled to two reservoirs
via one-channel leads. 
As we will show for such a scatterer
\begin{equation}
\label{Eq35}
 \hat A = 0,
\end{equation}

\noindent
and low frequency scattering up to linear in $\omega$ terms
is entirely described by the frozen scattering matrix $\hat S_{0}(t)$,
see Eqs.(\ref{Eq12}) and (\ref{Eq11}) at $\hat A=0$.

To find $\hat S_{F}$ we have to solve the Schr\"odinger equation
with the potential $V(x,t)$ being the delta function $\delta(x)$ multiplied 
by the amplitude oscillating in time:
\begin{equation}
\label{Eq36}
\begin{array}{l}
i\hbar\frac{\partial\Psi}{\partial t}   = \left( 
-\frac{\hbar^2}{2m_e}\frac{\partial^2}{\partial x^2}
+ V(x,t) \right) \Psi, \\
V(x,t) = \delta(x) \Big(V_0 + 2V_1\cos(\omega t + \varphi) \Big).
\end{array}
\end{equation}
According to the Floquet theorem 
the solution of the above equation 
has the form of Eq.(\ref{Eq3}). 
Away from the point $x=0$ the functions $\psi(E_n)$ are the plain waves: 
\begin{equation}
\label{Eq37}
\psi(E_n) = a_n e^{ik_n x} + b_n e^{-ik_n x}.
\end{equation}
The coefficients $a_n$, $b_n$ are determined from the boundary condition
at $x=0$:
\begin{equation}
\label{Eq38}
\begin{array}{c}
\Psi(x=+0) = \Psi(x=-0), \\
\ \\
\left.\frac{\partial\Psi}{\partial x}\right|_{x=+0} -
\left.\frac{\partial\Psi}{\partial x}\right|_{x=-0}   
= \frac{2m_e}{\hbar^2}V(t)\Psi(x=0).
\end{array}
\end{equation}
First, to find $S_{F,LL}$ and $S_{F,RL}$ 
we consider the plain wave of a unit amplitude with energy $E$ coming
from the left (we directed the $x$-axis from the left to the right):
\begin{equation}
\label{Eq39}
\Psi^{(in)}(E,t) = e^{-i\frac{E}{\hbar}t} e^{ikx}.
\end{equation}

\noindent Here $E=\hbar^2k^2/(2m_e)$.
Then the coefficients $a_n^{(out)}$ and $b_n^{(out)}$ for an outgoing wave
\begin{equation}
\label{Eq40}
\Psi^{(out)} = e^{-i\frac{E}{\hbar}t} \sum\limits_{n}
e^{-in\omega t} 
\left( \theta(x)a_n^{(out)}e^{ik_nx} +  \theta(-x)b_n^{(out)}e^{-ik_nx} \right),
\end{equation}

\noindent
(here $\theta(x)$ is the Heaviside step function: $\theta(x)=1$ at $x>0$ and
$\theta(x)=0$ at $x<0$)
define the Floquet scattering matrix elements as follows:
\begin{equation}
\label{Eq41}
\begin{array}{c}
S_{F,RL}(E_n,E) = \sqrt{\frac{k_n}{k}} a_n^{(out)}, \\
S_{F,LL}(E_n,E) = \sqrt{\frac{k_n}{k}} b_n^{(out)}.
\end{array}
\end{equation}
Substituting the whole wave function 
$\Psi = \Psi^{(in)} + \Psi^{(out)}$ 
into the boundary condition Eq.(\ref{Eq38})
we get the following relations between the different
$a_n^{(out)}$ and $b_n^{(out)}$:
\begin{equation}
\label{Eq42}
\begin{array}{c}
(k_n + i\kappa_0)a_{n}^{(out)} = k_n\delta_{n0} - 
i(\kappa_1a_{n-1}^{(out)} + \kappa_{-1}a_{n+1}^{(out)}), \\
\ \\
b_n^{(out)} = a_n^{(out)} - \delta_{n0}. 
\end{array}
\end{equation}

\noindent
Here we have introduced the following parameters:
\begin{equation}
\label{Eq43}
\kappa_0 = \frac{m_e}{\hbar^2}V_0, \quad
\kappa_{\pm 1}= \frac{m_e}{\hbar^2}V_1e^{\mp i\varphi}.
\end{equation}
We solve Eq.(\ref{Eq42}) in the adiabatic limit $\omega\to 0$ of interest here.
In this limit we can safely expand the wave vector $k_n$ as follows:
\begin{equation}
\label{Eq44}
k_n = k + \frac{n\omega}{v} + O(\omega^2),
\end{equation}

\noindent 
where $v=\hbar k/m_e$ is an electron velocity.
In addition we use the adiabatic expansion Eq.(\ref{Eq12A}) and express
$\hat S_{F}(E_n,E)$ in terms of the Fourier coefficients of the
matrix $\hat S_{in}(E)$. 
Substituting Eqs.(\ref{Eq41}), (\ref{Eq44}) and (\ref{Eq12A}) into 
Eq.(\ref{Eq42}), and ignoring all the terms of order $\omega^2$ and higher 
we can write:
$$
\begin{array}{c}
(k + i\kappa_0)S_{in,RL,n} + \left(\frac{1}{2} 
- i\frac{\kappa_0}{2k} \right)\frac{n\omega}{v}S_{0,RL,n} =  \\
k\delta_{n0} - i(\kappa_1 S_{in,RL,n-1} + \kappa_{-1}S_{in,RL,n+1}) \\
+ i\frac{\omega}{2vk}
[\kappa_1(n-1) S_{0,RL,n-1} + \kappa_{-1}(n+1)S_{0,RL,n+1}] , \\
\ \\
S_{in,LL,n}(E) = S_{in,RL,n}(E) - \delta_{n0}\sqrt{\frac{k_n}{k}}.
\end{array}
$$
Performing the inverse Fourier transformation we find
the equation for the time-dependent matrix elements of the matrix 
$\hat S_{in}(E,t)$:
\begin{equation}
\label{Eq45}
\begin{array}{c}
S_{in,RL}(E,t)  =  \frac{k}{k + i\kappa(t)} 
- \frac{i}{2vk} \frac{k - i\kappa(t)}{k + i\kappa(t)}
\frac{\partial S_{0,RL}(E,t)}{\partial t}, \\
\ \\
S_{in,LL}(E,t) = S_{in,RL}(E,t) - 1.
\end{array}
\end{equation}

\noindent
Here $\kappa(t) = m_eV(t)/\hbar^2$.
We solve these equations perturbatively in 
the small parameter proportional to $\partial/\partial t \sim \omega\to 0$.
To find the matrix elements $S_{in,RR}$ and $S_{in,LR}$ one can either 
exploit the symmetry condition or solve the same problem
but with the unit wave incoming from the right:
$\Psi^{(in)}(E,t) = e^{-i\frac{E}{\hbar}t} e^{-ikx}$.
Up to terms linear in $\omega$, the solution of both problems reads:
\begin{equation}
\label{Eq46}
\hat S_{in}(E,t) = \hat S_{0}(E,t) + \frac{i\hbar}{2}
\frac{\partial^2 \hat S_{0}(E,t)}{\partial t\partial E}.
\end{equation}

\noindent 
Here we used $\partial k/\partial E = 1/(\hbar v)$.
The stationary matrix is well known:
\begin{equation}
\label{Eq47}
\hat S_{0} = \frac{k}{k+i\kappa}\left(
\begin{array}{cc}
1 & 1 \\
1 & 1
\end{array}
\right) - \hat I.
\end{equation}

Comparing equations (\ref{Eq46}) and (\ref{Eq11A}) we arrive at 
the announced result, Eq.(\ref{Eq35}). 
Thus, to describe the low frequency scattering on point-like scatterer
it is enough to know only the frozen scattering matrix.

Alternatively, one can use Eq.(\ref{Eq9}) to show
that the matrix $\hat A$ vanishes for the oscillating $\delta$-function potential.
It is because the commutator ${\cal P}\{\hat S_{0}^{\dagger};\hat S_{0}\}$
is identically zero for the scattering matrix Eq.(\ref{Eq47}).
We can conclude that a point scatterer can not generate a quantum pump effect
since it can not rectify ac currents
(the spectral density $dI_{\alpha\beta}/dE$, Eq.(\ref{Eq27}), vanishes).
An oscillating scatterer does of course generate ac-currents, but these currents 
are total time derivatives of the charge near the barrier \cite{MBac03} and thus can not contribute to a dc-current.

Note, that the deviation of the effective scattering matrix $\hat S_{in}(E,t)$,
Eq.(\ref{Eq46}) 
from the frozen scattering matrix $\hat S_{0}(E,t)$, Eq.(\ref{Eq47}), 
is as small as, at least, $\hbar\omega/E$. 
For the opaque barrier the deviation is even smaller due to the factor
$k/\kappa\ll 1$. For the small oscillating amplitude case the deviation 
is additionally damped by the factor $\kappa_1/\kappa_0\ll 1$.

\subsection{Scatterer composed of two point-like barriers}
\label{SE2D}

In this subsection  we consider an example of a spatially "extended" scatterer 
which consists of two point-like scatterers 
placed at $x=0$ and $x=L$, respectively.
This system is coupled to two reservoirs via single channel leads.

The scattering properties of point scatterers are assumed to be 
oscillating in time with the same frequency $\omega$.
Scattering at the left and on the right barriers 
is described via the Floquet scattering matrices
$\hat S^L_{F}$ and $\hat S^R_{F}$, respectively. 
Scattering on the whole system is described via the Floquet scattering matrix
$\hat S_{F}$.

By analogy with the previous example we consider scattering of a unit wave
coming from the left, Eq.(\ref{Eq39}).
The whole wave function is of a Floquet function type Eq.(\ref{Eq3}) with
\begin{equation}
\label{Eq48}
\psi(E_n) = \left\{
\begin{array}{cc}
\delta_{n0}e^{ikx} + \sqrt{\frac{k}{k_n}}S_{F,LL}(E_n,E)e^{-ik_n x}, & x<0, \\
a_ne^{ik_nx} + b_ne^{-ik_nx}, & 0 < x < L, \\
\sqrt{\frac{k}{k_n}}S_{F,RL}(E_n,E)e^{ik_n (x-L)}, & x>L,
\end{array}
\right. .
\end{equation}
To find the unknown coefficients we use the boundary conditions
which we formulate in terms of scattering matrices 
$\hat S^L_{F}$ and $\hat S^R_{F}$ assumed to be known:
\begin{equation}
\label{Eq49}
\begin{array}{l}
S_{F,LL}(E_n,E) = S^L_{F,LL}(E_n,E) +
\sum\limits_{m} S^L_{F,LR}(E_n,E_m) \sqrt{\frac{k_m}{k_n}}b_m, \\
\frac{k_n}{k}a_n =  S^L_{F,RL}(E_n,E) +
\sum\limits_{m} S^L_{F,RR}(E_n,E_m)\sqrt{\frac{k_m}{k_n}} b_m, \\
\frac{k_n}{k}b_ne^{-ik_nL} = \sum\limits_{m}
S^R_{F,LL}(E_n,E_m) \sqrt{\frac{k_m}{k_n}}a_me^{ik_mL}, \\
S_{F,RL}(E_n,E) = \sum\limits_{m}
 S^R_{F,RL}(E_n,E_m) \sqrt{\frac{k_m}{k_n}}a_me^{ik_mL}.
\end{array}
\end{equation}
To simplify this system of equations we use the adiabatic approximation
Eq.(\ref{Eq12}) for the Floquet scattering matrices.
For this approximation to be valid, the energy quantum $\hbar\omega$
should be small compared with the relevant energy scale for the problem,
[see, Eq.(\ref{Eq9})].

In the case under consideration, there are several energy scales.
The first one is determined by the energy $E$ of an incoming electron. 
This scale relates to the deviation of the effective scattering matrices
$\hat S^L_{in}$ and $\hat S^R_{in}$  for point-like scatterers
from the corresponding frozen ones. This deviation is
of the order of $\hbar\omega/E$.
Another energy scale $\delta E$ relates to the spatial size of the system $L$
and arises from the quantum mechanical interference in the region between the 
scatterers at $0 < x < L$. 
In our case, Eq.(\ref{Eq49}), the interference effect is described 
via the factors $e^{ik_mL}$ which we will expand as follows:
\begin{equation}
\label{Eq50}
e^{\pm ik_mL} = e^{\pm ikL}\left( 1 \pm im\frac{\omega}{\omega_L} 
+ O(\omega^2) \right).
\end{equation}

\noindent 
Here $\omega_L = v/L$ defines the distance $\Delta E \sim \hbar\omega_L$
between the quantum levels if the system is decoupled from the reservoirs.
The second term in the brackets on the RHS of Eq. (\ref{Eq50}) is due to
an interplay of a quantum-mechanical interference with a quantized 
energy exchange between the scatterer and an electron traversing it.

The system can be treated as spatially "extended"  if $L\gg \lambda_E$, where 
$\lambda_E=h/\sqrt{2m_eE}$ is the de Broglie wave length for an electron with
energy E. In such a case the non-adiabatic corrections to 
the frozen scattering matrix are at least of order 
$\hbar\omega/\Delta E$.
Note if the energy $E$ is close to the energy of a transmission resonance then
the corrections will be of order $\hbar\omega/\Gamma$, where 
$\Gamma$ is the width of the transmission resonance.
In contrast, if  $L\ll \lambda_E$ then the scatterer can be viewed as point-like
and the non-adiabatic corrections will be as small as $\hbar\omega/E$
(see, Sec.\ref{SED}).
Therefore, assuming $L \gg \lambda_E$ we can safely ignore
the corrections of order $\hbar\omega/E$ and concentrate on 
the larger corrections of order 
$\hbar\omega/\delta E$ with $\delta E = \min\{\Delta E, \Gamma\}$.
Since we ignore the terms of order $\hbar\omega/E$ 
we can replace the Floquet scattering matrices for point-like scatterers
by the corresponding frozen scattering matrices, 
$\hat S^{R/L}_{F}(E_n,E_m) = \hat S^{R/L}_{0,n-m} + O(\hbar\omega/E)$
[see, Eq.(\ref{Eq8})]. 
To avoid a possible misunderstanding we do not write the energy $E$ as an
argument of $\hat S^{R/L}$ emphasizing that these matrices 
can be treated as energy independent on the scale of order $\delta E$.
Nevertheless they can depend on energy over a much larger scale, say, of order $E$. 
On the other hand, since we keep the terms of order $\hbar\omega/\delta E$
we use the adiabatic approximation 
$\hat S_{F}(E_n,E) = \hat S_{in,n}(E) + O(\omega^2)$  [see, Eq.(\ref{Eq12})] 
for the Floquet scattering matrix of the whole structure.

Using these approximations and substituting Eq.(\ref{Eq50}) into 
the system of equations (\ref{Eq49}) and performing 
the inverse Fourier transformation
we arrive at the following time-dependent equations 
valid up to first order in $\partial /\partial t$:
\begin{equation}
\label{Eq51}
\begin{array}{l}
S_{in,LL}(E,t) = \hat S^L_{0,LL}(t) + S^L_{0,LR}(t) b(t), \\
\ \\
a(t) = S^L_{0,RL}(t) + S^L_{0,RR}(t)\ b(t), \\
\ \\
e^{-ikL}\left(b(t) + \frac{1}{\omega_L} \frac{db(t)}{dt} \right) = 
S^R_{0,LL}(t)e^{ikL}\left(a(t) - \frac{1}{\omega_L} \frac{da(t)}{dt}\right), \\
\ \\
S_{in,RL}(E,t) = 
S^R_{0,RL}(t)e^{ikL}\left(a(t) - \frac{1}{\omega_L} \frac{da(t)}{dt}\right),
\end{array}
\end{equation}
Here we introduced the functions $a(t)$ and $b(t)$
defined as follows (x = a,b):
\begin{equation}
\label{Eq52}
x(t) = \sum\limits_{n} e^{-in\omega t} \sqrt{\frac{k_n}{k}}x_n.
\end{equation}
We consider the terms $da/dt$ and $db/dt$ as small perturbations and solve 
the system of equations (\ref{Eq51}) up to linear order in these corrections terms.

Note, that without the terms $da/dt$ and $db/dt$ the system of equations
Eq.(\ref{Eq51}) is exactly the system of equations which defines
the matrix elements of the frozen (stationary) scattering matrix 
(with the evident replacement $\hat S_{in}\to\hat S_{0}$).

Analogously, to calculate $S_{in,RR}$ and $S_{in,LR}$ we consider
the same problem but with the unit wave coming from the right:
$\Psi^{(in)}(E,t) = e^{-i\frac{E}{\hbar}t} e^{-ik(x-L)}$.
It is convenient to represent the results in the matrix form:
\begin{equation}
\label{Eq53}
\hat S_{in}(E,t) = \hat S_{0} - 
\frac{1}{\omega_L}\hat M_{L} \hat M^{-1} \frac{\partial }{\partial t}
\left[\hat M^{-1} \hat M_{R} \right].
\end{equation}

\noindent
Here $\bar S_0$ is the frozen scattering matrix:
\begin{equation}
\label{Eq54}
\hat S_{0}(E,t) = \hat M_0 + \hat M_{L} \hat M^{-1} \hat M_{R}. 
\end{equation}
The matrices $\hat M$ are all expressed in terms of the scattering matrix
elements for the left and right scatterers. 
They depend on energy through the factor $e^{ikL}$ and on time
through the matrices $\hat S^L_{0}$ and $\hat S^R_{0}$:
\begin{equation}
\label{Eq55}
\begin{array}{c}
\hat M_{0} = 
\left(
\begin{array}{cc}
S^L_{0,LL} & 0 \\
0 & S^R_{0,RR}
\end{array}
\right), 
\quad \hat M = 
\left(
\begin{array}{cc}
1 & - S^L_{0,RR} \\
 - S^R_{0,LL}e^{i2kL} & 1
\end{array}
\right), \\
\ \\
\hat M_{L} = 
\left(
\begin{array}{cc}
0 & S^L_{0,LR} \\
S^R_{0,RL}e^{ikL} & 0
\end{array}
\right), 
\quad \hat M_{R} = 
\left(
\begin{array}{cc}
S^L_{0,RL} & 0 \\
0 & S^R_{0,LR}e^{ikL}
\end{array}
\right). 
\end{array}
\end{equation}
Our aim is to calculate the matrix 
$$
\hbar\omega\hat A = \hat S_{in}-\hat S_{0}
-\frac{i\hbar}{2}\frac{\partial^2\hat S_{0}}{\partial t\partial E}
$$ 
[see, Eq.(\ref{Eq11A})]  for the double scatterer structure under consideration. 
Using Eq.(\ref{Eq54}) we obtain
$$
\frac{i\hbar}{2}\frac{\partial^2 \hat S_{0}}{\partial t\partial E} =
-\frac{1}{2\omega_L}\frac{\partial }{\partial t}
\left[\hat M_{L} \Big( \hat M^{-1}\Big)^2  \hat M_{R} \right].
$$
Then using Eq.(\ref{Eq53}) we obtain
\begin{equation}
\label{Eq56}
\begin{array}{c}
\hbar\omega\hat A(E,t) = \frac{1}{2\omega_L}
\left\{
\frac{\partial }{\partial t}
\left[\hat M_{L} \hat M^{-1} \right]\hat M^{-1}\hat M_{R} \right. \\
\ \\
\left.
- \hat M_{L} \hat M^{-1}\frac{\partial }{\partial t}
\left[\hat M^{-1}\hat M_{R}\right]
\right\}.
\end{array}
\end{equation}
The advantage of the above expression is its compactness.
However to be able to draw some conclusion we need the matrix
elements of $\hat A$ expressed directly in terms of the matrix elements
of  $S^L_{0}$ and $S^R_{0}$:
\begin{equation}
\label{Eq57}
\begin{array}{c}
\hbar\omega A_{LL} = \frac{S_{0,LL}-S^L_{0.LL}}{\omega_L\Delta}
\frac{\partial }{\partial t}
\left[ \ln\left(\frac{S^L_{0,LR}}{S^L_{0,RL}} \right)\right], \\
\ \\
\hbar\omega A_{LR} = \frac{S_{0,LR}}{2\omega_L\Delta}
\left\{(2-\Delta)\frac{\partial }{\partial t}
\left[ \ln\left(\frac{S^L_{0,LR}}{S^R_{0,LR}} \right)\right] \right. \\
\left. + (1-\Delta)\frac{\partial }{\partial t}
\left[ \ln\left(\frac{S^R_{0,LL}}{S^L_{0,RR}} \right)\right]
\right\}, \\
\ \\
\hbar\omega A_{RL} = -\frac{S_{0,RL}}{2\omega_L\Delta}
\left\{(2-\Delta)\frac{\partial }{\partial t}
\left[ \ln\left(\frac{S^L_{0,RL}}{S^R_{0,RL}} \right)\right] \right. \\
\left. + (1-\Delta)\frac{\partial }{\partial t}
\left[ \ln\left(\frac{S^R_{0,LL}}{S^L_{0,RR}} \right)\right]
\right\}, \\
\ \\
\hbar\omega A_{RR} = \frac{S_{0,RR}-S^R_{0.RR}}{\omega_L\Delta}
\frac{\partial }{\partial t}
\left[ \ln\left(\frac{S^R_{0,RL}}{S^R_{0,LR}} \right)\right].
\end{array}
\end{equation}

\noindent 
Here $\Delta= 1 - S^L_{0,RR}S^R_{0,LL}e^{i2kL}$.
We see that for the case considered
the matrix elements of $\hat A$ are proportional to time derivatives
as it should be.

We can now use Eq.(\ref{Eq57}) to investigate the symmetry properties 
of the matrix $\hat A$.
We suppose that the matrices $S^L_{0}$ and $S^R_{0}$ are of the form 
given by Eq.(\ref{Eq33}). 
If there is no magnetic field, $H=0$, then in Eq.(\ref{Eq33}) 
the phases $\phi^{L/R} = 0 $. 
Therefore, the scattering matrices are symmetric in the lead indices,
$S^{L/R}_{\alpha\beta} = S^{L/R}_{\beta\alpha}$.
In such a case $A_{\alpha\alpha} = 0$ and $A_{LR} = - A_{RL}$
that is in agreement with Eq.(\ref{Eq23}).
In addition, if the system is inversion symmetric,
i.e., if the scatterers are the same, $\hat S^L = \hat S^R$,
and if in addition $\theta=0$ and $\phi=0$,
then all matrix elements are zero, $\hat A = 0$.
Note, for an oscillating scatterer the last two properties apply only 
if two scatterers oscillate in synchronism, but not if they oscillate 
with a phase lag.

\subsection{Ring with enclosed magnetic flux}
\label{SER}

\begin{figure}[t]
  \vspace{0mm}
  \centerline{
   \epsfxsize 8cm
   \epsffile{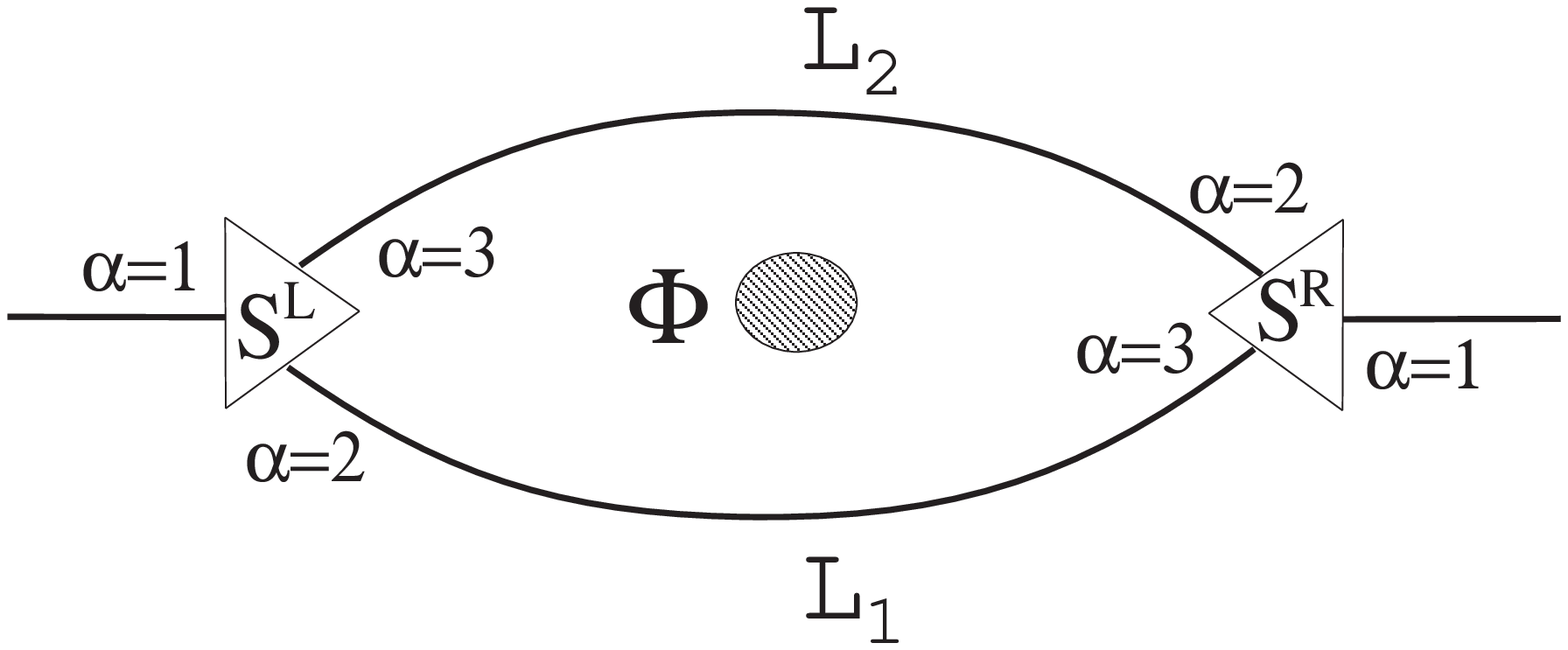}
             }
  \vspace{0mm}
  \nopagebreak
  \caption{ A one-channel ring of length $L = L_1 + L_2$
with enclosed magnetic flux $\Phi$
and with two leads. The Greek letter $\alpha$ numbers the scattering channels
at the left $S^L$ and right $S^R$ wave splitters. 
}
\label{fig1}
\end{figure}

Our third example is a one-channel ring 
with enclosed magnetic flux $\Phi$ coupled to two reservoirs, Fig.\ref{fig1}.
The lower and the upper branches of the ring have length $L_1$ and $L_2$,
respectively.  
Following Ref.~\onlinecite{BIA84}
we describe the coupling between the ring and the lead via
a single parameter $3\times 3$ scattering matrix 
$\hat S_{0}(\epsilon)$, 
with $\epsilon = \epsilon^L$ and $\epsilon = \epsilon^R$ 
for the left and the right coupling points, respectively.
The numbering of scattering channels is shown in Fig.\ref{fig1}.
We will use two different matrices, 
$\hat S_{0}^{(s)}$ and  $\hat S_{0}^{(a)}$, which
couple the lead to the branches of the ring symmetrically and 
asymmetrically, respectively. 
They are:
\begin{subequations}
\label{Eq58}
\begin{equation}
\label{Eq58A}
\hat S_{0}^{(s)}(\epsilon) = \left(
  \begin{array}{ccc}
       -(a+b) & \sqrt{\epsilon} & \sqrt{\epsilon} \\
       \sqrt{\epsilon}  &  a  &  b  \\
       \sqrt{\epsilon}  &  b  &  a         
   \end{array}
\right),
\end{equation}
\begin{equation}
\label{Eq58B}
\hat S_{0}^{(a)}(\epsilon) = \left(
  \begin{array}{ccc}
       a  &  \sqrt{\epsilon}  &   b  \\
       \sqrt{\epsilon} & -(a+b) &  \sqrt{\epsilon} \\
        b & \sqrt{\epsilon}    &  a         
   \end{array}
\right).
\end{equation}
\end{subequations}

\noindent
Here
$a = (\sqrt{1-2\epsilon} - 1)/2$ and
$b = (\sqrt{1-2\epsilon} + 1)/2$.
The coupling parameter should be within the following interval:
$0\leq \epsilon\leq 0.5$. 
We suppose that the coupling parameters $\epsilon^L$ and $\epsilon^R$
oscillate in time with the same frequency $\omega$
but with the phase lag $\Delta\varphi = \varphi^R - \varphi^L$:
$$
\epsilon^{L/R} = [\epsilon^{L/R}_0 
+ \epsilon^{L/R}_1\cos(\omega t + \varphi^{L/R})]/2. 
$$
We keep the parameters $\epsilon^L$ and $\epsilon^R$
independent of the electron energy. 
This allows us to describe the time-dependent scattering at the three lead splitters 
within the frozen scattering matrix approximation, see Sec.\ref{GAAA0}. 
Also we assume that a voltage $V$ can be applied between the 
reservoirs, here kept at the same temperature, see Eq.(\ref{Eq32}).

\begin{figure}[t]
  \vspace{0mm}
  \centerline{
   \epsfxsize 8cm
   \epsffile{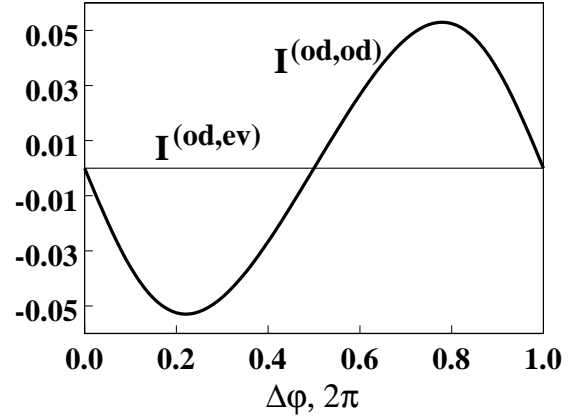}
             }
  \vspace{0mm}
  \nopagebreak
  \caption{ 
The parts of the pumped currents which are odd in magnetic flux
$I_{L}^{(od,ev)}$ and $I_{L}^{(od,od)}$, 
Eqs.(\ref{Eq34D}) and (\ref{Eq34E}), 
are given as a function of the phase difference 
$\Delta\varphi = \varphi^R - \varphi^L$
between the oscillating coupling parameters $\epsilon^R(t)$ and $\epsilon^L(t)$ for
symmetric coupling:
Both the left and the right wave splitters are coupled symmetrically
to the branches of the ring. 
The currents are in units of $e\omega/(2\pi)$.
The parameters are:
$L_1=\pi$; $L_2=0.7\pi$; $\Phi=0.4\Phi_0$; $\mu_0=9$;
$V = 0.5$; $T_L = T_R = 0.1$;
$\epsilon^L_{0} = \epsilon^R_{0}=0.5$;
$\epsilon^L_{1}= \epsilon^R_{1}=0.4$.
We use the following units: $2m_e=1$; $\hbar=1$; $e=1$; $k_B=1$.
}
\label{fig2}
\end{figure}

\begin{figure}[b]
  \vspace{0mm}
  \centerline{
   \epsfxsize 8cm
   \epsffile{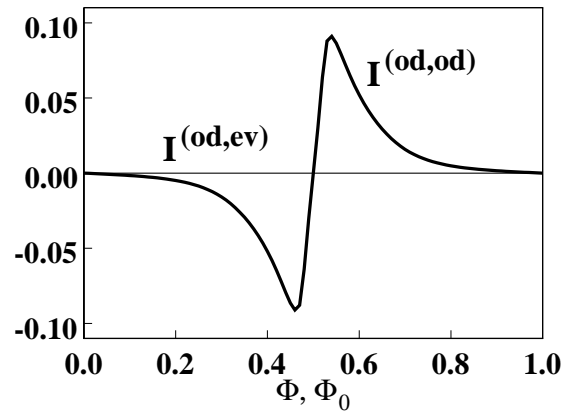}
             }
  \vspace{0mm}
  \nopagebreak
  \caption{ 
The currents $I_{L}^{(od,ev)}$ and $I_{L}^{(od,od)}$
[in units of $e\omega/(2\pi)$]
as a function of the enclosed magnetic flux $\Phi$ [in units
of the magnetic flux quantum $\Phi_0 = h/e$] for the case of 
symmetric coupling:
Both the left and the right wave splitters are coupled symmetrically
to the branches of a ring.
The parameters are the same as in Fig.\ref{fig2}.
The phase difference is: $\Delta\varphi = \pi/2.$
}
\label{fig3}
\end{figure}

We concentrate mainly on the part of the current $I_{L}^{(od)}$
which is odd in magnetic flux.
This current consists itself of two parts, 
$I_{L}^{(od,ev)}$ and $I_{L}^{(od,od)}$, 
which are defined by  Eqs.(\ref{Eq34D}) and (\ref{Eq34E}), respectively.
To find $I_{L}^{(od,od)}$ it is necessary to calculate the residual Floquet
matrix $\hat A$ for the system under consideration, Fig.\ref{fig1}.
These calculations are quite similar to those given in Sec.\ref{SE2D}
and we do not give the details here, see Appendix.

The current
$I_{L}^{(od,ev)}$, 
odd in magnetic flux and even in applied voltage, 
is governed by the phase $\phi$ which determines 
the asymmetry in the transmission phase through the ring, see Eq.(\ref{Eq33}).
Interestingly, in our model this asymmetry depends crucially
on the type of coupling between the leads and the ring.
If each of the leads is coupled symmetrically to the arms of the ring, 
i.e., if $\hat S^L_{0} = \hat S_{0}^{(s)}(\epsilon^L)$ 
and    $\hat S^R_{0} = \hat S_{0}^{(s)}(\epsilon^R)$,
then the phase $\phi = 0$ for any magnetic flux.
In such a case the current $I_{L}^{(od,ev)}$ is identically zero
and the full current odd in magnetic flux $I_{L}^{(od)}$ is odd in the applied
voltage as well. 

In Fig.\ref{fig2} and Fig.\ref{fig3} we depict the current contribution
which is odd in magnetic flux
as a function of the phase lag and the magnetic flux, respectively, 
for symmetric coupling.
Thus if the coupling between the ring and the leads is symmetric 
the current odd in magnetic flux appears only at 
$V\neq 0$ (and/or at $\Delta T\neq 0$), 
i.e., when the electron flows incident on the ring from the reservoirs 
are different.
In contrast, if any lead (or both) is coupled asymmetrically to the ring,
i.e., if $\hat S^L_{0} = \hat S_{0}^{(a)}(\epsilon^L)$ 
and/or    $\hat S^R_{0} = \hat S_{0}^{(a)}(\epsilon^R)$,
then the phase $\phi$ is not identically zero 
and the current $I_{L}^{(od,ev)}$ does contribute to $I_{L}^{(od)}$. 

\begin{figure}[t]
  \vspace{0mm}
  \centerline{
   \epsfxsize 8cm
   \epsffile{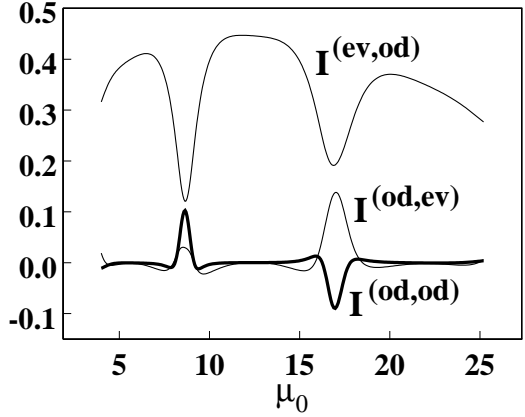}
             }
  \vspace{0mm}
  \nopagebreak
  \caption{ 
The currents through the ring are given as a function of the Fermi energy 
$\mu_{0}$.
The coupling is asymmetric:
Both the left and the right wave splitters 
are coupled asymmetrically to the branches of a ring.
The parameters are the same as in Fig.\ref{fig2}.
The phase difference is: $\Delta\varphi = \pi/2.$
The currents $I_{L}^{(od,ev)}$ and $I_{L}^{(od,od)}$ are given 
in units of $e\omega/(2\pi)$.
The current even in magnetic flux $I_{L}^{(ev,od)}$ is given 
in dimensionless units of  $1/(2\pi)$.
}
\label{fig4}
\end{figure}

In Fig.\ref{fig4} we depict the currents $I_{L}^{(od,ev)}$ and
$I_{L}^{(od,ev)}$ as a function of the Fermi energy $\mu_0$
for asymmetric coupling.
For comparison we give the current $I_{L}^{(ev,od)}$, Eq.(\ref{Eq34C}),
which is proportional to the transmission probability through the ring.
Both currents $I_{L}^{(od,ev)}$ and $I_{L}^{(od,od)}$
peak (by modulo) at energies where the transmission (reflection)
coefficient changes sharply. 

\section{Conclusion}
\label{C}

In this work we analyze the scattering properties of a periodically driven
mesoscopic scatterer. 
Traversing such a scatterer an electron can gain or lose one or several 
energy quanta $\hbar\omega$ and thus can change its energy.
Therefore, generally the scattering matrix of a periodically driven mesoscopic
scatterer depends on two energies, incoming and outgoing, 
We show that at low driving frequency $\omega\to 0$ one can introduce 
effective matrices depending on only one energy, 
either incoming or outgoing [see, Eq.(\ref{Eq11})], 
which approximates accurately the Floquet scattering matrix 
up to terms of order $\omega$ [see, Eq.(\ref{Eq12})].
We introduce two effective matrices, $\hat S_{in}$ and $\hat S_{out}$, 
which are not unitary. Nevertheless each of them conserves the current after 
averaging over a driving cycle.

The matrices $\hat S_{in}$ and $\hat S_{out}$ are the sum of a frozen
scattering matrix and a matrix which determines the linear in $\omega$ part. 
The last is responsible for the quantum pump effect \cite{Brouwer98}
and it consists of two contributions. The first one is the second derivative
of the frozen scattering matrix $\hat S_{0}(t)$.
The second contribution is defined by an in principle new matrix $\hat A$.
In particular, the matrix $\hat A$ has a symmetry with respect to magnetic
field reversal, Eq.(\ref{Eq23}), that is opposite to that of the stationary 
(frozen)
scattering matrix, Eq.(\ref{Eq17}). 
In contrast to the stationary scattering matrix 
the residual Floquet matrix reflects directly the chirality 
of the pumping process.

Using the adiabatic representation Eq.(\ref{Eq12}) for the Floquet scattering 
matrix we examine the dc current flowing through the 
two terminal (many channels) mesoscopic sample
under the simultaneous action of a slow parametric oscillation of the scatterer 
and simultaneously subject to an applied dc voltage.
We divide the current into parts with definite symmetry properties
with respect to a magnetic field and/or a voltage inversion.

As it is known in the stationary case the dc current 
through the coherent two terminal sample
is an even function of a magnetic field.
On the other hand the periodically driven scatterer shows an odd in magnetic
field, linear in $\omega$ current, Eq.(\ref{Eq31B}),
which is due to the quantum pump effect.
The odd in applied voltage part of this current is proportional to the residual 
Floquet matrix $\hat A$ [see also, Eq.(\ref{Eq34E})].

We demonstrate that the calculation of the residual matrix $\hat A$ can be performed 
in close analogy with the calculation of the stationary scattering matrix $\hat S_{0}$.
We emphasize that the matrix $\hat A$ reflects the interplay of 
absorbing/emitting of energy quanta $\hbar\omega$ with quantum mechanical
interference inside the scatterer. 
For instance, for a point-like scatterer 
(without the space for interference inside)
the matrix $\hat A$ is identically zero. 

Our work suggests that additional experiments which investigate 
a driven mesoscopic conductor 
in a less symmetric setup, i.e., with reservoirs having different
electrochemical potentials or temperatures, might be useful 
to reveal the presence of a quantum pump effect. 

\begin{acknowledgments}
We thank M.L. Polianski for a critical reading of the manuscript. 
This work was supported by the Swiss National Science
Foundation.
\end{acknowledgments}

\newpage 

\appendix*\section{}

\subsection*{Ring with enclosed magnetic flux: Analytical expressions} 
\label{A1} 

The stationary scattering matrix $\hat S_0$ for the ring with
branches of length $L_1$ and $L_2$, and with enclosed magnetic flux $\Phi$ 
coupled to two leads via wave splitters $S^L$ and $S^R$ (see, Fig.\ref{fig1})
reads:
$$
\begin{array}{c}
\hat S_{0} = \left(
\begin{array}{cc}
S_{0,LL} & S_{0,LR} \\
S_{0,RL} & S_{0,RR}
\end{array}
\right), \\
\ \\
S_{0,LL} = S^L_{11} + S^L_{12}W_2 
+ S^L_{13}W_3e^{iL_2\left(\frac{2\pi}{L}\frac{\Phi}{\Phi_0} + k\right)}, \\
\ \\
S_{0,RL} = S^R_{12}W_4 
+ S^R_{13}W_1e^{iL_1\left(\frac{2\pi}{L}\frac{\Phi}{\Phi_0} + k\right)}, \\
\ \\
S_{0,LR} = S^L_{12}W^{\prime}_4 
+ S^L_{13}W^{\prime}_1e^{iL_2\left(\frac{2\pi}{L}\frac{\Phi}{\Phi_0} 
+ k\right)}, \\
\ \\
S_{0,RR} = S^R_{11} + S^R_{12}W^{\prime}_2 
+ S^R_{13}W^{\prime}_3e^{iL_1\left(\frac{2\pi}{L}\frac{\Phi}{\Phi_0} 
+ k\right)}.
\end{array}
$$
\noindent
Here $L=L_1+L_2$; $\Phi_0=h/e$ is the single-electron magnetic flux quantum.
The vector-columns $\hat W$ and $\hat W^{\prime}$ are defined in the 
following way:
$$
\begin{array}{c}
\hat W = \hat M^{-1}\hat Y, \\
\ \\
\hat W^{\prime} = \hat M^{\prime-1}\hat Y^{\prime}, \\
\ \\
\hat Y = \left(
\begin{array}{c}
-S^L_{21} \\ -S^L_{31} \\ 0 \\ 0 
\end{array}
\right), \quad
\hat Y^{\prime} = \left(
\begin{array}{c}
-S^R_{21} \\ -S^R_{31} \\ 0 \\ 0 
\end{array}
\right), \\
\ \\
\hat M = \left(
\begin{array}{cccc}
-1 & S^L_{22} & S^L_{23}c[L_2,k] &  0\\ 
 0 & S^L_{32} & S^L_{33}c[L_2,k] & -c[L_2,-k] \\ 
S^R_{23}c[L_1,k] & 0 & -1 &  S^R_{22} \\ 
S^R_{33}c[L_1,k] & -c[L_1,-k] & 0 & S^R_{32}
\end{array}
\right), \\
\ \\
\hat M^{\prime} = \left(
\begin{array}{cccc}
-1 & S^R_{22} & S^R_{23}c[L_1,k] &  0\\ 
 0 & S^R_{32} & S^R_{33}c[L_1,k] & -c[L_1,-k] \\ 
S^L_{23}c[L_2,k] & 0 & -1 &  S^L_{22} \\ 
S^L_{33}c[L_2,k] & -c[L_2,-k] & 0 & S^L_{32}
\end{array}
\right). 
\end{array}
$$
Here we introduced the functions
$c[x,y] = e^{ix\left(\frac{2\pi}{L}\frac{\Phi}{\Phi_0} +y\right)}$.
The anti-symmetric matrix $\hat A$ 
characterizing the ability of the ring with oscillating coupling
parameters $\epsilon^{L}(t)$ and $\epsilon^{R}(t)$
[see, Eq.(\ref{Eq58})] to work as a pump is:
$$
\begin{array}{c}
\hbar\omega \hat A = \hat {\rm a} 
-\frac{i\hbar}{2}\frac{\partial^2\hat S_{0}}{\partial t\partial E}, \\
\ \\
\hat {\rm a} = \left(
\begin{array}{cc}
{\rm a}_{LL} & {\rm a}_{LR} \\
{\rm a}_{RL} & {\rm a}_{RR}
\end{array}
\right), \\
\ \\
{\rm a}_{LL} = S^L_{12}\delta W_2 
+ S^L_{13}
\left( \delta W_3 - \frac{1}{\omega_{L_2}}\frac{\partial W_3}{\partial t} \right)
e^{iL_2\left(\frac{2\pi}{L}\frac{\Phi}{\Phi_0} + k\right)}, \\
\ \\
{\rm a}_{RL} = S^R_{12}\delta W_4 
+ S^R_{13}
\left(\delta W_1 - \frac{1}{\omega_{L_1}}\frac{\partial W_1}{\partial t} \right)
e^{iL_1\left(\frac{2\pi}{L}\frac{\Phi}{\Phi_0} + k\right)}, \\
\ \\
{\rm a}_{LR} = S^L_{12}\delta W^{\prime}_4 
+ S^L_{13} \left(\delta W^{\prime}_1 
- \frac{1}{\omega_{L_2}}\frac{\partial W^{\prime}_1}{\partial t} \right)
e^{iL_2\left(\frac{2\pi}{L}\frac{\Phi}{\Phi_0} + k\right)}, \\
\ \\
{\rm a}_{RR} = S^R_{12}\delta W^{\prime}_2 
+ S^R_{13} \left( \delta W^{\prime}_3
- \frac{1}{\omega_{L_1}}\frac{\partial W^{\prime}_3}{\partial t} \right)
e^{iL_1\left(\frac{2\pi}{L}\frac{\Phi}{\Phi_0} + k\right)}.
\end{array}
$$
Here $\omega_{L_j} = v/L_i,$ j = 1,2.
The vector-columns $\delta\hat W$ and $\delta\hat W^{\prime}$ are:
$$
\begin{array}{c}
\delta\hat W = -i\hbar \hat M^{-1}\frac{\partial \hat M}{\partial E} 
\frac{\partial \hat W}{\partial t}, \\
\ \\
\delta\hat W^{\prime} = -i\hbar \hat M^{\prime-1}
\frac{\partial \hat M^{\prime}}{\partial E} 
\frac{\partial \hat W^{\prime}}{\partial t}.
\end{array}
$$
Note that all the quantities depend on energy $E=\hbar^2k^2/(2m)$ 
through the phase factors
$e^{\pm iL_jk}$, $j=1,2$ and on time $t$ through the scattering matrices of 
wave splitters $\hat S^{\beta}(t) = \hat S^{(s/a)}(\epsilon^{\beta}(t))$,
$\beta = L, R$ [see, Eqs.(\ref{Eq58})].

\end{document}